\documentclass[aps,prd,preprint,showpacs,amsmath,amssymb]{revtex4}
\usepackage{epsfig}
\newlength{\www}

\newcommand{\be}{\begin{equation}}
\newcommand{\ee}{\end{equation}}
\newcommand{\ba}{\begin{eqnarray}}
\newcommand{\ea}{\end{eqnarray}}

\newcommand{\bq}{\begin{equation}}
\newcommand{\eq}{\end{equation}}
\newcommand{\bqa}{\begin{eqnarray}}
\newcommand{\eqa}{\end{eqnarray}}
\newcommand{\ben}{\begin{enumerate}}
\newcommand{\een}{\end{enumerate}}
\newcommand{\bc}{\begin{center}}
\newcommand{\ec}{\end{center}}
\newcommand{\bqb}{\begin{eqnarray*}}
\newcommand{\eqb}{\end{eqnarray*}}

\newcommand{\qsl}{q\hskip-0.21cm\slash}
\newcommand{\qpsl}{q'\hskip-0.29cm\slash}
\newcommand{\esl}{e\hskip-0.21cm\slash}
\newcommand{\eesl}{\epsilon\hskip-0.21cm\slash}

\begin{document}

\draft
\preprint{PTA/06-2}

\title{\vspace{1cm}
Associated $\lowercase{t}W$ production at LHC:\\
a complete calculation of electroweak supersymmetric\\
effects at one loop
}

\author{M. Beccaria$^{a,b}$, G. Macorini$^{c, d}$
F.M. Renard$^e$ and C. Verzegnassi$^{c, d}$ \\
\vspace{0.4cm}
}

\affiliation{\small
$^a$Dipartimento di Fisica, Universit\`a di
Lecce \\
Via Arnesano, 73100 Lecce, Italy.\\
\vspace{0.2cm}
$^b$INFN, Sezione di Lecce\\
\vspace{0.2cm}
$^c$
Dipartimento di Fisica Teorica, Universit\`a di Trieste, \\
Strada Costiera
 14, Miramare (Trieste) \\
\vspace{0.2cm}
$^d$ INFN, Sezione di Trieste\\
$^e$ Laboratoire de Physique Th\'{e}orique et Astroparticules,
UMR 5207\\
Universit\'{e} Montpellier II,
 F-34095 Montpellier Cedex 5.\hspace{2.2cm}\\
\vspace{0.2cm}
}

\begin{abstract}
We compute, in the MSSM framework, the total electroweak contributions at
one loop for the process $pp\to tW+X$, initiated by the parton process $bg\to tW$. The 
supersymmetric effect is analyzed for various choices of the SUSY benchmark points.
Choosing realistic unpolarized and polarized experimental quantities, 
we show the size of the various effects and discuss their dependence
on the MSSM parameters.
\end{abstract}


\maketitle

\section{Introduction}
\label{sec:intro}

The relevance of the process of associated $tW$ production from proton - proton collisions has been exhaustively
stressed in recent dedicated studies~\cite{Yellow}. In the Standard Model framework, it is well known that
accurate measurements of the production rate would provide an excellent determination of the $tbW$ coupling.
For physics beyond the SM, one expects that precision tests of virtual effects might be performed, provided that the effects
were {\em sufficiently} large, i.e. at least of the same size as the overall theoretical and experimental uncertainty.
On the theoretical side, an estimate given in~\cite{Yellow} predicts for the total production cross section an uncertainty of
about 15\%. On the experimental side one must recall the fact that the considered process will be seen, for the first time, at LHC, 
simply because of the required pp energy. Therefore an estimate of the expected experimental uncertainty is in fact still missing.
This might be particularly relevant, if the estimated effect turned out to be {\em reasonable} (e.g. of the same size as the theoretical
one), for the special purpose of performing a precision test of supersymmetric models, in particular
of the simplest available one, the MSSM.
In fact, in a previous paper~\cite{Beccaria:2004xk}, the genuinely weak effects of the model were considered at one loop for all
the processes of single top production ($td$, $t\overline{b}$, $tW$ and also $tH^-$) at LHC. The treatment
was rather preliminary and qualitative, and only considered the very special case of a light SUSY scenario
and of a production in a large ($\sim 1$ TeV) final invariant mass range, where a simple logarithmic expansion of so called
Sudakov kind could be used. The feature that emerged was that, for the three processes that will certainly be seen
at LHC (i.e. $td$, $t\overline{b}$, $tW$), the electroweak relative effect in the MSSM is sizable,
particularly for large $\tan\beta$ values where it could reach the 30 \% size. This 
appeared to us a good motivation for performing a complete accurate estimate, valid for all
realistic invariant masses and containing all the parameters of the model. In this paper, 
we present the results of our effort for the $tW$ production process. It is, to our knowledge, the first complete calculation
of the electroweak MSSM effects at one loop for the process, that also includes the QED soft photon
radiation. We have checked the validity of our results in three different ways, i.e. 
(a) we have verified the cancellation of all the (virtual) ultraviolet divergences, (b) we have verified the cancellation
of all the (real and virtual) infrared divergences and, last but not least, (c) we have verified the exact reproduction
of the asymptotic Sudakov expansions, given in~\cite{Beccaria:2004xk}, from the computed Feynman diagrams. After
these three checks we hope that our results should be correct, and we shall show them in the paper with the 
following plan: in Section~\ref{sec:diagrams} a brief description of the relevant Feynman diagrams is given, and a discussion of the cancellation
of the ultraviolet and infrared divergences is provided. 
Section~\ref{sec:sudakov} shows the reproduction of the (essentially academic) asymptotic Sudakov expansion.
In Section~\ref{sec:effects}, the realistic observables are defined, and the related MSSM effect is shown for various choices of the SUSY
benchmark points.
A final discussion that included a review of future calculations is provided in Section V. 

\section{MSSM $bg\to \lowercase{t}W$ production at one loop}
\label{sec:diagrams}

The process that we have considered is the so called exclusive associated $tW$ production, whose partonic 
description corresponds to the two body final state reaction
\be
bg\to tW^-
\ee
that is represented, at Born level, in Fig.~(1). In this paper we shall
not consider the inclusive process 
\be
gg\to t\overline{b}W^- .
\ee
As known~\cite{bdist}, the collinear $\overline{b}$ component
of this process is already enclosed as a QCD NLO correction to the bottom quark
distribution function of the exclusive process Eq.~(1) and our treatment will only consider the one-loop electroweak effects.

At the one loop level, we have to consider different kinds of Feynman diagrams, several of which will exhibit an 
ultraviolet divergence. We shall  choose the {\em on-shell} renormalization scheme, and in this framework we shall define
the following classes of Feynman diagrams:

\subsection{Born, self-energies and counterterms}

The two Born diagrams represented in Fig. 1 (a, b) are an $s$-channel b quark exchange and a $u$-channel top
quark exchange. With the definitions
$s=(p_b+p_g)^2=(p_W+p_t)^2$, and $u=(p_b-p_W)^2=(p_g-p_t)^2$ channel, we have
\bqa
A^{Born}(gb\to Wt)={eg_s\over s_W\sqrt{2}}
\bar u(t)[{\eesl P_L(\qsl+m_b)\esl \over s-m^2_b}
+{\esl (\qpsl+m_t)\eesl P_L\over u-m^2_t}]u(b)
\eqa
\noindent
where
\bq
q=p_g+p_b=p_W+p_t~~~s=q^2~~~~~q'=p_t-p_g=p_b-p_W~~~u=q^{'2},
\eq
and $e$, $\epsilon$ are the gluon and $W$ polarization vectors, respectively.

In the on-shell renormalization scheme, these Born terms have to be completed with counterterms
associated to the b, t, and W lines. These counterterms are expressed in terms of quark and gauge bosons
self energy functions $\Sigma^f_{L, R, S}(k^2)$, $\Sigma^{VV'}(k^2)$.
In these self-energies we take into account SM and SUSY contributions (sfermions, Higgs, neutralinos and charginos).
The b and t quark propagators are also modified by self-energy functions of $s$ and $u$.

In the $s$-channel, we can use the invariant forms 
\bq
I^s_{1L,R}=\eesl\qsl\esl P_{L,R}~~~~
I^s_{2L,R}=\eesl\esl P_{L,R}
\eq
and write the amplitude as 
\bq
A=\sum_{\eta}\{N^{s~\eta}_1I^s_{1\eta}+N^{s~\eta}_2I^s_{2\eta}
\}
\eq
where
\bqa
N^{s~L}_1&=&{eg_s\over\sqrt{2}s_W(s-m^2_b)}\{
1+\delta Z^W_1-\delta Z^W_2+{1\over2}\delta\Psi_W
+{1\over2}\delta\Psi_t\nonumber\\
&&+{3\over2}\delta Z^b_L
+{1\over2}\delta Z^t_L
-~{s\over s-m^2_b}(\Sigma^b_L(s)+\delta Z^b_L)\nonumber\\
&&
-~{m^2_b\over s-m^2_b}(\Sigma^b_R(s)+\delta Z^b_R)
-~{2m^2_b\over s-m^2_b}[\Sigma^b_S(s)
-{1\over2}(\delta Z^b_L+\delta Z^b_R)-{\delta m_b\over m_b}]\}
\eqa
\bq
N^{s~R}_1=0~~~~N^{s~L}_2=0
\eq
\bqa
N^{s~R}_2&=&{eg_sm_b\over\sqrt{2}s_W(s-m^2_b)}\{
1+\delta Z^W_1-\delta Z^W_2+{1\over2}\delta\Psi_W
+{1\over2}\delta\Psi_t\nonumber\\
&&+{1\over2}\delta Z^b_L
+{1\over2}\delta Z^t_L+\delta Z^b_R
-~{s\over s-m^2_b}(\Sigma^b_R(s)+\delta Z^b_R
+\Sigma^b_L(s)+\delta Z^b_L)\nonumber\\
&&
-~{s+m^2_b\over s-m^2_b}[\Sigma^b_S(s)
-{1\over2}(\delta Z^b_L+\delta Z^b_R)-{\delta m_b\over m_b}]\}
\eqa
In the $u$-channel, we define 
\bqa
I^u_{1L,R}&=&\esl\qsl'\eesl P_{L,R}~~~~
I^u_{2L,R}=\esl\eesl P_{L,R}
\eqa
and write 
\bq
A=\sum_{\eta}\{N^{u~\eta}_1I^u_{1\eta}+N^{u~\eta}_2I^u_{2\eta}
\}
\eq
with 
\bqa
N^{u~L}_1&=&{eg_s\over\sqrt{2}s_W(u-m^2_t)}\{
1+\delta Z^W_1-\delta Z^W_2+{1\over2}\delta\Psi_W
+{1\over2}\delta\Psi_t\nonumber\\
&&+{3\over2}\delta Z^t_L
+{1\over2}\delta Z^b_L
-~{u\over u-m^2_t}(\Sigma^t_L(u)+\delta Z^t_L)\nonumber\\
&&
-~{m^2_t\over u-m^2_t}(\Sigma^t_R(u)+\delta Z^t_R)
-~{2m^2_t\over u-m^2_t}[\Sigma^t_S(u)
-{1\over2}(\delta Z^t_L+\delta Z^t_R)-{\delta m_t\over m_t}]\}
\eqa
\bq
N^{u~R}_1=0~~~~N^{u~R}_2=0
\eq
\bqa
N^{u~L}_2&=&{eg_sm_t\over\sqrt{2}s_W(u-m^2_t)}\{
1+\delta Z^W_1-\delta Z^W_2+{1\over2}\delta\Psi_W
+{1\over2}\delta\Psi_t\nonumber\\
&&+{1\over2}\delta Z^b_L
+{1\over2}\delta Z^t_L+\delta Z^t_R
-~{u\over u-m^2_t}(\Sigma^t_R(u)+\delta Z^t_R
+\Sigma^t_L(u)+\delta Z^t_L)\nonumber\\
&&
-~{u+m^2_t\over u-m^2_t}[\Sigma^t_S(u)
-{1\over2}(\delta Z^t_L+\delta Z^t_R)-{\delta m_t\over m_t}]\}
\eqa
The various counterterms have the following explicit expressions in terms of self-energies.
First, we have the divergent quark wave function renormalizations
\bq
\delta Z^b_L=\delta Z^t_L\equiv \delta Z_L=
-\Sigma^b_L(m^2_b)-m^2_b[\Sigma^{'b}_L(m^2_b)+\Sigma^{'b}_R(m^2_b)
+2\Sigma^{'b}_S(m^2_b)]
\eq
\bq
\delta Z^b_R=
-\Sigma^b_R(m^2_b)-m^2_b[\Sigma^{'b}_L(m^2_b)+\Sigma^{'b}_R(m^2_b)
+2\Sigma^{'b}_S(m^2_b)]
\eq
\bq
\delta Z^t_R=\delta Z_L+\Sigma^t_L(m^2_t)-\Sigma^t_R(m^2_t)
\eq
Then, we have the finite wave-function renormalization required in the on-shell scheme
and unavoidable since we have both up and down type quarks in the process
\bq
\delta\Psi_t=-\{\Sigma^t_L(m^2_t)+\delta Z_L+
m^2_t[\Sigma^{'t}_L(m^2_t)+\Sigma^{'t}_R(m^2_t)
+2\Sigma^{'t}_S(m^2_t)]\}
\eq
The similar terms for the $W$ gauge boson are

\bq
\delta Z^W_1-\delta Z^W_2=~{\Sigma^{\gamma Z}(0)\over s_Wc_W M^2_Z}
\eq
\bq
\delta Z^W_2 = - \Sigma^{'\gamma\gamma}(0) 
+2{c_W\over s_W M^2_Z}\Sigma^{\gamma Z}(0) 
+{c^2_W\over s^2_W}[{\delta M^2_Z\over M^2_Z} - 
{\delta M^2_W\over M^2_W}]
\eq
and
\bq
\delta\Psi_W=-\Sigma^{'WW}(M^2_W)=
-\{\Sigma^{'WW}(M^2_W) + \delta Z^W_2\}
\eq
Finally, we list the mass counterterms
\bq
\delta M^2_W=Re\Sigma^{WW}(M^2_W)~~~~~\delta M^2_Z=Re\Sigma^{ZZ}(M^2_Z)
\eq
\bq
\delta m_b={m_b\over2}Re[\Sigma^b_L(m^2_b)+\Sigma^b_R(m^2_b)
+2\Sigma^b_S(m^2_b)]
\eq
\bq
\delta m_t={m_t\over2}Re[\Sigma^t_L(m^2_t)+\Sigma^t_R(m^2_t)
+2\Sigma^t_S(m^2_t)]
\eq

\subsection{Vertex corrections and Box diagrams}

The next two classes of diagrams are triangle-like vertices and box diagrams. A list of the generic
diagrams (i.e. diagrams with virtual particles left unspecified apart from their spin) is shown
in Fig.~(\ref{fig:generic}) as produced by FeynArts~\cite{FeynArts}. Schematically we can further subdivide them as follows
($q$ stands for b or t quarks, V for $\gamma, Z, W$, $H$ for neutral or charged Higgses or Goldstone particles,
$\chi$ for chargino or neutralino):

\begin{enumerate}
\item {\em Initial} $s$-channel triangles connected to the intermediate $b$ quark:  $(Vqq)$, $(Hqq)$, $(\chi \tilde q \tilde q$);
\item {\em Final} $s$-channel triangles connected to the intermediate $b$ quark: $(btV)$, $(HHq)$, $(\tilde b\tilde t \chi)$
$(VVq)$, $(HVq)$, $(VHq)$, $(btH)$, $(\chi\chi\tilde q)$;
\item {\em Up} $u$-channel triangles connected to the intermediate $t$ quark: $(qqV)$, $(qqH)$, $(\tilde q\tilde q\chi)$, 
\item {\em Down} $u$-channel triangles connected to the intermediate $t$ quark: $(tbV)$, $(tbH)$, $(\chi\chi\tilde q)$, $(VVq)$, $(VHq)$, $(HVq)$, $(HHq)$, 
$(\tilde t\tilde b\chi)$;

\item {\em Direct} boxes: $(\tilde b \tilde b \tilde t \chi^0)$, $(b b t V)$, $(b b t H)$;

\item {\em Crossed} boxes: $(qqVV)$, $(qqVH)$, $(qqHV)$, $(qqHH)$, $(\tilde q\tilde q\chi\chi)$;

\item {\em Twisted} boxes: $(ttVb)$, $(tt Hb)$,  $(\tilde t\tilde t\chi^0 \tilde b)$.
\end{enumerate}

The notation corresponds to the clockwise ordering of the internal particles inside the diagrams. 

An essential step consists then in checking the cancellation of UV divergences. They appear in the self energy functions
$\Sigma(k^2)$, in the various counterterms and in the various triangles. Box contributions are convergent. We have 
checked the cancellation when summing all of these terms. This cancellation
occurs in several independent sectors (gauge, Higgs, SM, SUSY).

Having completed the first important check (cancellation of UV divergences) we now move to the forthcoming issue of cancellation
of IR divergences that will be treated in the forthcoming discussion.

\subsection{Cancellation of IR divergences}

QED radiation effects are usually split into a soft part containing the potential IR singular terms, and a 
hard part including the emission of photons with energy not small compared to the process energy scale.
In this brief section, we discuss the soft emission and the detailed cancellation of IR divergences
that occurs when it is combined with virtual photon exchanges.

Let us denote by ${\cal A}^{Born}$ and ${\cal A}^{1 loop}$ any invariant helicity scattering amplitude
evaluated at Born or one loop level. Let us also denote by $\lambda$ the photon mass acting as an IR regulator.
The IR cancellation between (soft) real radiation and virtual photon exchange 
holds in every helicity channel separately and we have checked it numerically.

It reads
\be
\left({\cal A}^{Born}\right)^2 \left(1 + \frac{\alpha}{2\pi}\delta_s\right) + 2 {\cal A}^{Born}\ {\cal A}^{1 loop} = \mbox{finite as}\ \lambda\to 0
\ee
where, in the above expressions, $\delta_S$ is the correction factor taking into account the emission of soft real 
photons with energy from $\lambda$ up to $E_\gamma^{max} << \sqrt{s}$. The explicit expression for $\delta_S$ can be found, 
for instance, in~\cite{'tHooft:1978xw}.

In practice the above relation follows from 
the eikonal factorization
\be
\label{eq:eikonal}
{\cal A}^{1 loop} =- {\cal A}^{Born} \frac{\alpha}{4\pi}\delta_s  + \mbox{regular terms as}\ \lambda\to 0
\ee
It is possible to split further the above factorization property. Indeed, the singular part of the radiation factor 
has the form 
\be
\delta_S = \log\frac{\lambda}{E_\gamma^{max}}\ \sum_{i,j} \delta_S^{i,j}   + \mbox{regular terms as}\ \lambda\to 0
\ee
where $i$ and $j$ runs over the initial/final charged particles, i.e. (bt), (bW), and (tW). 
There are two types of contributions:
the diagonal ones with $i=j$ and the off diagonal ones with $i\neq j$~\cite{'tHooft:1978xw}.

Now, the matching between the singular $\log\lambda$ in the l.h.s. and r.h.s. of Eq.~(\ref{eq:eikonal})
can be checked in several independent steps as follows
\begin{enumerate}

\item the diagonal radiation terms $i=j$ match the IR divergence in the counterterms
associated to the $i$-th external line~\cite{Weinberg:1995mt}.

\item the off-diagonal radiation terms $i\neq j$ match the IR divergence in the diagrams
which are obtained connecting in all ways the $i$-th and $j$-th external lines with a 
virtual photon. This operation produces both triangle and box diagrams.
\end{enumerate}

As a final comment, we remark that gauge invariance is crucial to cancel all 
non factoring contributions associated to the final $W$ line as discussed in~\cite{Lemoine:1979pm}.

The next step in the treatment of QED effect is the calculation of hard photon emission. We have left
this subject to a dedicated study which shall be discussed separately~\cite{Pavia}.

\section{Sudakov expansion of the scattering amplitudes}
\label{sec:sudakov}

Let us now consider the high energy behavior of the $bg\to tW$ helicity 
amplitudes $F_{\lambda\mu\lambda'\mu'}$, where $\lambda,\mu,\lambda',\mu'$
refer to the helicities $\lambda_b,\lambda_g,\lambda_t,\lambda_W$ respectively.
Several simplifications appear in the Born and in the one-loop contributions. 
When $s\gg m_i^2$ ($m_i$ being the internal or external 
involved masses), ignoring $m_i^2/s$ contributions, the non-suppressed Born 
amplitudes reduce to $F_{----}$, $F_{-+-+}$ for 
transverse $W$ and $F_{-++0}$ for longitudinal $W$.

The leading high energy Born helicity amplitudes are

\be
F^{Born}_{----}\to {eg_s\over s_W\sqrt{2}}({\lambda^l\over2})
{2\over \cos{\theta\over2}}
\ee

\be
F^{Born}_{-+-+}\to {eg_s\over s_W\sqrt{2}}({\lambda^l\over2})
{2 \cos{\theta\over2}}
\ee

\be
F^{Born}_{-++0}\to {eg_s\over s_W}({\lambda^l\over2})
{m_t\over M_W}
\cos{\theta\over2}({1-\cos\theta\over1+\cos\theta})
\label{fbornl}\ee

Note that $F_{-++0}$
is controlled by the top Yukawa coupling factor $\sim m_t/M_W$.  In fact
the amplitude $F_{+--0}$ also occurs but at a much weaker level as it is 
controlled by the bottom Yukawa coupling factor $\sim m_b/M_W$.

At one loop, these amplitudes receive logarithmic enhancements as discussed
in several papers, called Sudakov terms. These
terms are separated into universal and into angular dependent components. 
From the rules established in~\cite{NewReality},
one expects the following
expressions (there are misprints in the paper~\cite{Beccaria:2004xk}; The correct equations
are the following Eqs.~(31-46)).

For transverse $W$ amplitudes:

\ba
F^{Univ}_{-,\mu,-,\mu}&=&
F^{Born}_{-,\mu,-,\mu}
[~{1\over2}~(~c^{ew}(b\bar b)_{L}
+c^{ew}(t\bar t)_{L}~)~+c^{ew}(W_T)]
\ea
\be
c^{ew}(q\bar q)_L=c^{ew}(\tilde{q}\tilde{\bar q})_L=
c(q\bar q, ~gauge)_L~+~c(q\bar q,~yuk)_L
\ee

\be
c(d\bar d, ~gauge)_L=c(u\bar u, ~gauge)_L={\alpha(1+26c^2_W)\over144\pi
s^2_Wc^2_W}~(n~ \log{s\over m^2_W}-\log^2{s\over m^2_W})
\ee

\be
c(d\bar d, ~gauge)_R={\alpha\over36\pi c^2_W}
~(n~ \log{s\over m^2_W}-\log^2{s\over m^2_W})
\ee

\be
c(u\bar u, ~gauge)_R={\alpha\over9\pi c^2_W}
~(n~ \log{s\over m^2_W}-\log^2{s\over m^2_W})
\ee

where $n=3,2$ in SM and MSSM, respectively.

\be
c(b\bar b, ~yuk)_L=
c(t\bar t, ~yuk)_L=-~{\alpha\over16\pi s^2_W}~
[\log{s\over m^2_W}]~
[{m^2_t\over m^2_W}y_t+{m^2_b\over m^2_W}y_b]
\ee
\be
c(b\bar b, ~yuk)_R=-~{\alpha\over8\pi s^2_W}~
[\log{s\over m^2_W}]~
[{m^2_b\over m^2_W}y_b]
\ee
\be
c(t\bar t, ~yuk)_R=-~{\alpha\over8\pi s^2_W}~
[\log{s\over m^2_W}]~
[{m^2_t\over m^2_W}y_t]
\ee

where $y_t=1,~2(1+\cot^2\beta)$ and $y_b=1,~2(1+\tan^2\beta)$ in SM and MSSM, respectively.

\be
c^{ew}(W_T)={\alpha\over4\pi s^2_W}[-\log^2\frac{s}{M_W^2}]
\ee

and for  the longitudinal $W^-_0$ amplitude:

\ba
F^{Univ}_{-,+,+,0}&=&
F^{Born}_{-,+,+,0}
[~{1\over2}~(~c^{ew}(b\bar b)_{L}
+c^{ew}(t\bar t)_{R}~)~+c^{ew}(W_0)]
\label{funivl}
\ea
with, in SM:

\ba
c^{ew}(W_0)&=&{\alpha\over4\pi}\{~
[-~{1+2c^2_W\over8s^2_Wc^2_W}\log^2\frac{s}{M_W^2}]\nonumber\\
&&+[\log\frac{s}{M_W^2}][-~{15-42 c^2_W\over72s^2_Wc^2_W}+
{3(m^2_t-m^2_b)\over8s^2_WM^2_W}]~\}
\ea
such that
\ba
F^{Univ}_{-,+,+,0}&=&
F^{Born}_{-,+,+,0}[~{\alpha\over4\pi}]\{~[-\log^2\frac{s}{M_W^2}]
[{13+14c^2_W\over36s^2_Wc^2_W}]\nonumber\\
&&
+[{1+2c^2_W\over2s^2_Wc^2_W}-{m^2_b\over2s^2_Wc^2_W}]
[\log\frac{s}{M_W^2}]~\}
\label{funivlsm}
\ea

whereas in MSSM :

\ba
c^{ew}(W_0)&=&{\alpha\over4\pi}\{
[-~{1+2c^2_W\over8s^2_Wc^2_W}\log^2\frac{s}{M_W^2}]\nonumber\\
&&+[\log\frac{s}{M_W^2}][-~{17+10 c^2_W\over36s^2_Wc^2_W}+
{m^2_b\over4s^2_WM^2_W}(1+\tan^2\beta)+
{3m^2_t\over4s^2_WM^2_W}(1+\cot^2\beta)]\}\nonumber\\
&&
\ea
such that
\ba
F^{Univ}_{-,+,+,0}&=&
F^{Born}_{-,+,+,0}[~{\alpha\over4\pi}]\{~[-\log^2\frac{s}{M_W^2}]
[{13+14c^2_W\over36s^2_Wc^2_W}]~\}
\label{funivlsm}
\ea
\noindent
(in which all single logs cancel !).

For the  electroweak angular terms we find:

\ba
F^{ang}_{-,\mu,-,\mu}&=&
F^{Born}_{-,\mu,-,\mu}[-~{\alpha\over2\pi}][\log\frac{s}{M_W^2}]\{~[\log{-t\over s}]
[{1-10c^2_W\over36s^2_Wc^2_W}]
+~{1\over s^2_W}\log{-u\over s}~\}
\label{fangt}\ea
\ba
F^{ang}_{-,+,+,0}&=&
F^{Born}_{-,+,+,0}[-~{\alpha\over24\pi c^2_W}][\log\frac{s}{M_W^2}]
\{~[\frac{4}{3} \log{-t\over s}]-{1-10c^2_W\over s^2_W}\log{-u\over s}~\}
\label{fangl}\ea

Note that the longitudinal $W$ amplitudes satisfy the equivalence theorem which states that, neglecting $m_i^2/s$ contributions, they 
should coincide with the amplitudes for the process $bg\to tG^-$, $G^-$ being the charged Goldstone boson.

We have checked, by using the asymptotic expansions of the B, C, D functions appearing in the self-energies, triangle and
box amplitudes that our full one-loop result produces the logarithmic contributions expected by the rules
given above.

These resulting asymptotic expressions deserve several comments. In the case of transverse $W$ production, one checks that 
at Born and one-loop level and at next-to-leading logarithmic accuracy in addition to trivial fermion 
chirality constraint $\lambda_t = \lambda_b = -1/2$ gauge boson helicity conservation~\cite{Gounaris:2005ey} is preserved, both in SM
and MSSM cases, i.e. only $\mu = \mu'$ amplitudes survive. One then sees that the MSSM differs from the SM in the single 
logarithm contributions, $n=2$ instead of $n=3$ for gauge terms and $2(1+\cot^2\beta)$ or $2(1+\tan^2\beta)$ Yukawa enhancements, 
especially large for large $\tan\beta$.

In the case of longitudinal $W$ production, the Born amplitude is controlled by the Yukawa $m_t/M_W$ factor
associated to fermion chirality violation $\lambda_t = -\lambda_b = 1/2$ and satisfies also the rule 
$\lambda_g + \lambda_b = \lambda_t$ which is an extension of the GBHC rule~\cite{Gounaris:2005ey}. An additional remarkable feature
appears for the single log contribution, namely it totally cancels in the MSSM case.

Having successfully performed the ultraviolet, infrared and Sudakov {\em tests}, we
hope that our complete expressions will be correct. In this respect, we should
add the following comment: We do not expect that, at lower energies and for
higher SUSY masses, the simple features that we met in the {\em light SUSY}
Sudakov description given in~\cite{Beccaria:2004xk} retain their full validity. Still, we
would expect that, at least, some of the main features could survive. For
instance, for what concerns the slope of the invariant mass distribution, we
could hope that a simple modification at lower energies, or at lower
energy/SUSY masses ratios, might be the addition of a (possibly {\em large})
constant term at least in a {\em moderate} energy region not too far from the
asymptotic one, so that a smooth connection between the two regions is
achieved. In the following section we shall return on this point, but first we
shall define and examine those quantities that will be the realistic
experimental observables.

\section{Physical Predictions}
\label{sec:effects}

We are now able to provide numerical predictions for the complete electroweak 
effect of the MSSM at one loop on the realistic observables of the considered 
tW production process. 
With this aim, we shall divide our presentation in two
parts, that correspond respectively to the consideration of unpolarized and of 
polarized quantities. Following a pragmatic attitude i.e. assuming that only 
unpolarized observables will be measured in a first stage of the experiments, 
we shall start our analysis with the former ones.
       
The first quantity that we shall consider is the invariant mass distribution,
conventionally defined as
\ba
\label{eq:basic}
{d\sigma(PP\to t W^-+X)\over ds}&=&
{1\over S}~\int^{\cos\theta_{max}}_{\cos\theta_{min}}
d\cos\theta~[L_{bg}(\tau, \cos\theta)
{d\sigma_{bg\to  t W^-}\over d\cos\theta}(s)~]
\ea
\noindent
where $\tau={s\over S}$, and $L_{bg}$ is the parton process luminosity.

\be
L_{bg}(\tau, \cos\theta)=
\int^{\bar y_{max}}_{\bar y_{min}}d\bar y~ 
~[~ b(x) g({\tau\over x})+g(x)b({\tau\over x})~]
\label{Lij}
\ee
\noindent
where S is the total pp c.m. energy, and 
$i(x)$ the distributions of the parton $i$ inside the proton
with a momentum fraction,
$x={\sqrt{s\over S}}~e^{\bar y}$, related to the rapidity
$\bar y$ of the $tY$ system~\cite{QCDcoll}.
The parton distribution functions are the latest NNLO MRST (Martin, Roberts, Stirling, Thorne) 
set available on~\cite{lumi}.
The limits of integrations for $\bar y$ depends on the cuts. We have chosen a 
maximal rapidity $Y=2$ and a minimum $p_T$ which we shall specify later.

Note that we are at this stage considering as kinematical observable the 
initial partons c.m. energy $\sqrt{s}$, and not the realistic final 
state invariant mass $M_{tW}$. The transition from the first quantity to the 
second one can be performed using the available suitable event 
generators, like for instance PYTHIA~\cite{Pythia}, as we did in a previous paper on top-antitop production~\cite{Beccaria:2004sx}. 
We expect from that experience a small (few percent) modification in 
the transition from $\sqrt{s}$ to $M_{tW}$. This correction can be considered as a {\em QCD 
effect}, and as such it will be consistently treated in a forthcoming paper~\cite{forth1}
where this type of non electroweak effects will be included.
For what concerns the complete one-loop electroweak amplitude, we can compute it 
for any choice of the MSSM parameters, but before doing this 
we want to show some features of the simple Born approximation of the partonic 
amplitude that we consider particularly relevant for an understanding of our following results. 
More precisely, the point that we want to stress is that the partonic invariant 
scattering amplitude for the process, that represents the starting block of 
our calculations, turns out to be the sum of twelve different helicity 
amplitudes, that have been defined already in Section~\ref{sec:sudakov}. 
For large values of $\sqrt{s}$, i.e. for $\sqrt{s}$  sufficiently larger than the masses of 
all the particles and sparticles involved in the one-loop description of the 
process, we expect that only three helicity amplitudes remain dominant, more 
precisely those that have been defined in Section~\ref{sec:sudakov} as $F_{----}$, 
$F_{-+-+}$, $F_{-++0}$ (the third and fourth index specifies the top and W helicity).
The remaining helicity amplitudes vanish asymptotically i.e. for $s\to\infty$ like $1/s$ with 
possible logarithmic corrections at one loop, and in our preliminary paper~\cite{Beccaria:2004xk}
they  were systematically neglected in the region of that was considered, 
corresponding to a) energies in the 1 TeV range and b) {\em light} SUSY masses 
scenario. For the realistic analysis that we can now carry on, both assumptions 
will be abandoned. In particular:

a) The possibility of identifying the final $(t,W)$ signal must face the serious 
competition of a background, mostly due to events coming from the copious 
top-antitop and WWj production.This problem has been already exhaustively 
discussed in a previous paper~\cite{bdist}, where it 
has been shown that the introduction of suitable b-tagging cuts will allow to 
extract the signal at reasonable (20 fb${}^{-1}$) luminosities. A priori, one would 
expect that the background contamination should be under control for c.m. 
energies below a qualitative {\em background threshold} of approximately, say, 
400-500 GeV, and increase in the higher energies region. Keeping this 
limitation in mind, we have nonetheless analyzed in this paper the full energy 
region from threshold to 1 TeV, although at this final energy value the 
identification of the signal might be difficult. The reason of this 
(optimistic) choice is that we do not have yet at disposal a  rigorous experimental analysis of 
the realistically expected size of the signal at variable energies, as we had 
in the preliminary top-antitop paper~\cite{Beccaria:2004sx}. This analysis 
is being already performed, and will be included in the already mentioned 
forthcoming work.

b) The SUSY scenario that we shall investigate is the conventional mSUGRA one.
In particular, we shall consider a number of benchmark points that are nowadays available, 
trying to choose those that show a
definite difference in the values of the various SUSY masses, and of $\tan\beta$.
We  insist on the fact that we could perform its calculation for 
any choice of the parameters, but for obvious reasons we have limited the 
presentation of Figures in this paper. 

After these preliminary remarks, we now show in Fig.~(\ref{fig:diffsigmaborn})
the comparison (treated in Born approximation) of  quantities 
that we consider particularly worth of being considered, i.e. the parton c.m. 
angular dependences of the differential cross section in various helicity channels.
We have chosen four c.m. energy values, $\sqrt{s} = 300$, 500, 1000, 
and (academically) 2000 GeV and retained for sake of comparison the full angular 
range $-1\le\cos\theta\le 1$ (possible angular cuts 
will be considered separately). We have only retained those terms that are 
numerically meaningful, leaving aside the {\em invisible} ones. In the Figure, for simplicity,
we show only the 5 amplitudes which are leading at high energy. These are the three asymptotic ones 
generated by the helicity amplitudes previously defined and two extra ones, 
corresponding to $F_{---0}$ and $F_{-++-}$.
The important points to be noticed are the 
following ones:

\begin{enumerate}
\item The relative relevance of the different helicity differential cross sections 
changes drastically with the scattering angle for the two lower energy 
points. As one sees, the scattering  in the nearly backward region is totally 
dominated for $\sqrt{s} = 300-500$ GeV by the two {\em non asymptotic} quantities; the weight  of the 
asymptotic differential cross sections becomes  dominant when $\theta$ moves to 
the forward direction, where the overall numerical size is, though, smaller 
than that of the backward region.

\item Although less evidently, these features survive also at the next energy 
point $\sqrt{s} = $ 1 TeV. More precisely, the size of the $F_{---0}$ distribution remains 
essential in the backward region.

\item One might start doubting about the validity of our asymptotic assumptions. 
To show that this is not the case, we have plotted the distributions in the 
last sub-figure, for the (academic) point $\sqrt{s} = 2$ TeV. As one sees, the features 
at this energy are those that would expect at (sufficiently!!) high 
energies:the largely dominant contribution is that of two of the asymptotic 
quantities, more precisely $F_{-++0}$ and $F_{----}$.
\end{enumerate}

In conclusion, we see that the contribution of the {\em non asymptotic} helicity 
amplitudes, for which no Sudakov expansion has to be expected, is essential 
for realistic (i.e., qualitatively  $<$ 1 TeV) energies. A {\em proper} asymptotic 
behavior seems to eventually set in, but only at higher energies (say, $\sim 2$ 
TeV), where the possibility of detecting the signal appears, least to 
say, debatable.Although these features were derived by an analysis performed in 
Born approximation, we expect that the complete results that will follow will 
be consistent with these preliminary impressions.

\section{Results}
\label{sec:results}

The successful results of our previous tests  have encouraged us to prepare 
with a reasonable amount of confidence a numerical C++ code that 
contains the complete {\em tested} one-loop expression of all the components of 
the considered process. This program has been called MINSTREL and is nowadays 
working and available. Thanks to this code we 
are now able to provide numerical predictions for the complete electroweak 
effect of the MSSM at one loop on the realistic observables of the associated $tW^-$
production.

With this aim, we have returned to Eq.~(\ref{eq:basic}) and have considered a set of SUSY 
benchmark points that appeared to us suited for our analysis. More precisely, 
we have retained representative points whose SUSY masses values are not 
{\em light} ( but not even dramatically large) and also points whose masses are, conversely, 
{\em light} (in our language, lighter than, say, 400-500 GeV). Also, 
we have used  points whose only essential difference is the value of $\tan\beta$, 
that is allowed to become definitely large (50) in one of the two cases and 
still appreciable (10) in the second one. In this way, we should be able to 
compare the complete results with those that we found in the {\em light SUSY}
Sudakov approximation. For practical reasons, we will only show the results of 
our analysis for a choice of four representative points. Two of them are the 
ATLAS DC2 SU1 and SU6 points~\cite{DC2}; the remaining two are two points 
whose spectrum has been evaluated by the code SUSPECT~\cite{Djouadi:2002ze} and that 
we have called LS1, LS2 where LS stands for Light SUSY.
To make 
the reasons of our choice evident, we have given in Tab.~(\ref{tab:spectra}) the values of the 
various SUSY masses, and of $\tan\beta$, that correspond to the four choices.
One 
sees that the first two points correspond to a {\em not light} choice, with two 
different values of $\tan\beta$; for the last two points, a {\em light} SUSY scenario is 
assumed, with, again, two different $\tan\beta$ values.
Two final technical points have to be now added: 

\begin{enumerate}
\item[a)] Our calculations have been performed with a value of $p_{T, \min} = 15$ GeV. 
\item[b)] In the calculations, we have included a QED soft photon contribution, 
computed assuming an upper value of the soft photon energy $\Delta E = 0.1$ GeV.
As we anticipated, the full treatment of the essentially Standard Model hard photon 
emission will be contained in a dedicated paper~\cite{Pavia}.
\end{enumerate}

\subsection{Unpolarized observables}

\subsubsection{Effects in the distribution $d\sigma/ds$}

We can now show the first results of our calculations. In Fig.~\ref{fig:sigma}  we have drawn 
the relative effect at one loop of the MSSM, and also of the SM alone, for the 
four choices of benchmark points. The calculation stops at $\sqrt{s}$ = 1 TeV as we announced.
From a glance at the different Figures, a number of 
(preliminary) conclusions can already be drawn. In particular:

\begin{enumerate}

\item[a)]
The {\em genuine} SUSY effect, i.e. the difference between the MSSM and the SM, 
remains systematically small (a relative few percent) for all choices of the 
benchmark points in the considered (realistic) energy region. In this sense, a 
measurement of the invariant mass distribution of the process does not appear 
to be a promising way of detecting genuine SUSY effects in the MSSM with 
mSUGRA symmetry breaking (this conclusion could be not valid for 
different supersymmetric models or symmetry breaking scenarios).

\item[b)] The relative effect of the considered MSSM is not, though, negligible. As 
one sees, it varies from positive to negative values in the lowest part of the 
region, remaining systematically negative for larger energies and reaching a 
common value of approximately ten percent around 1 TeV. This large energy negative shift 
from the Born level calculation appears a characteristic property of the 
considered MSSM model, independent of the values of the parameters that were 
assumed in our analysis.
\end{enumerate}

\subsubsection{Ratios of partially integrated cross sections: a proposal}

Since there is a wide energy region where the one loop effects are appreciable, i.e. $\sqrt{s} \gtrsim 500$ GeV,
we can split it in two parts, compute the associated integrated cross section, and evaluate the 
ratio $R$ of the two partial cross sections. This investigation is motivated by the 
following remarks concerning general properties of $R$:
\begin{enumerate}
\item[a)] It should be free of several systematic experimental errors;
\item[b)] It should be free of several QCD effects (same pdfs, same virtual 
corrections);
\item[c)] It should be essentially unaffected by photon radiation effects.
\end{enumerate}

To give an explicit numerical example, we have considered the scenario SU6, and have split the 
high energy region in two parts:
\be
E_{\rm threshold} \equiv m_t + m_W\ < \sqrt{s} < 400\ \mbox{GeV},\qquad 
\sqrt{s} > 400\ \mbox{GeV}.
\ee
We call $\sigma_-$ and $\sigma_+$ the integrated cross section $\int( d\sigma/ds)\ ds$ in the two regions, and
define $R = \sigma_+/\sigma_-$.

We denote by $\varepsilon_\pm$ the relative MSSM effect on the two cross sections. We also denote by $N_\pm$
the expected number of events associated to the two regions. Of course, $N_\pm = {\cal L}\ \sigma_\pm$, where
${\cal L}$ is the luminosity. If we call $\Delta_{\rm MSSM} R$ and $\Delta_{\rm stat} R$ the MSSM and statistical shifts on $R$
we have 
\be
\Delta_{\rm MSSM}R = R(\varepsilon_+-\varepsilon_-),\qquad\Delta_{\rm stat}R = R\left(\frac{1}{\sqrt{N_+}} + \frac{1}{\sqrt{N_-}}\right).
\ee
In our test case, the Born value is $R\simeq 0.58$ and the difference $\varepsilon_+-\varepsilon_-$ gives a shift
$0.57\to 0.60$ of about 3.5 \%. the \underline{purely} statistical error computed with a luminosity 
${\cal L} = 10\ \mbox{fb}^{-1}$ gives $\Delta_{\rm stat} R\simeq 0.002$, i.e. a shift about 10 times smaller than the 
MSSM effect.

We conclude that radiative effects in ratios like $R$ are beyond the statistical noise. 
Of course, systematic errors are expected to dominate over statistical ones. Thus, a detailed
dedicated experimental study of the process reconstruction will be crucial to assess $R$ 
as a {\em realistic} observable and a potential precision test of the electroweak sector of the considered MSSM.

\subsubsection{Sudakov-like parameterizations}

To conclude the unpolarized session, we have tried to give an effective parametrization 
of the full one loop effect in the spirit of the logarithmic Sudakov expansion. 
As we remarked, a straightforward comparison with the results described in Sec.~\ref{sec:sudakov} 
is hampered by a variety of problems, that we now emphasize:

\begin{enumerate}
\item[a)] Box diagrams are functions of the Mandelstam invariants $t, u$ beside $s$. At small or 
large angles, these can be small (compared to the internal squared masses and $s$) and spoil the 
validity of the Sudakov approximation. 

\item[b)] At high but moderate energies (below 1 TeV) there are several subleading helicity channels which are 
relevant and non negligible. These channels certainly admit a Sudakov expansion. However this is not as simple
as that of the leading channels. The coefficients of the expansion for these amplitudes have not
been investigated before and could or could not turn out to be simple combination of 
quantum numbers and couplings as in the leading case.

\item[c)] In the MSSM, we have sparticles with masses around 300-400 GeV, even in the lightest considered scenario LS2.
The extent to which they can be regarded as {\em small} can only be determined by an explicit 
numerical comparison of the two calculations. Indeed, by a careful inspection of the various involved diagrams
one sees that box diagrams can display a rather delayed asymptotic behavior. In practice, if the typical virtual masses
are of order $m$, there are box diagrams with asymptotic behavior $\sim \log(\sqrt{s}/m')$ where $m'$ can be 
4-5 times larger than $m$, depending in particular on the scattering angle. This large effective scale contributes a large energy independent constant shift in the 
difference between the Sudakov and the one loop calculations. Also, since we always require $\sqrt{s}\gg m'$, it pushes 
forward the energy range where the expansion is accurate.

\end{enumerate}

As a consequence of these remarks, the difference between the full one-loop MSSM effect and the Sudakov approximation 
is expected to be a small, slowly varying function of the energy, at least in the considered energy range. On the contrary, 
in the Standard Model, 
all masses are quite light compared to the typical 500-1000 GeV parton energy and we can hope to observe a better 
accuracy of the Sudakov expansion.

All these expectations are confirmed by actual calculations. As an illustration, we show in Fig.~(\ref{fig:sudakov})
the comparison between the full one loop and the Sudakov calculations of the effect in the distribution $d\sigma/ds$.
The left panel shows the Standard Model case. The right panel shows the LS2 MSSM scenario, which is the lightest considered.
For purpose of comparison, we have switched off QED radiation and set $M_\gamma = M_Z$. 
We have computed the effects up to unrealistic values (2 TeV) of the energy, just to emphasize the 
convergence at high energy. 
The Sudakov approximation is evaluated with a common scale $\widetilde M$ in the double and single logarithms. The best value of $\widetilde M$
is an important issue and will be discussed below. The main features of the figure are the following.

\begin{enumerate}
\item[a)] In the Standard Model, we choose $\widetilde M = M_W$. We observe a remarkable agreement. The expansion is rather accurate down to energies
$\sqrt{s}\simeq 500 GeV$. The relevant scale is the electroweak breaking one $\simeq M_W$ and there are no large
constant (energy independent) contributions. 

\item[b)] In the LS2-MSSM case, we adjust $\widetilde M$ in order to have the same slope in the two curves. 
We have found the optimal value $\widetilde M = 120$ GeV.
With this choice, there is a large but constant shift of about +6\% with respect to the 
Sudakov calculation as shown in the upper right panel where we show the two curves. 
To emphasize the energy independence of the shift, we show in the lower right panel the same one loop curve together with a shifted 
version of the Sudakov one which has been moved upward by a constant +6\%. The agreement is again remarkable,
exactly like in the Standard Model case.
\end{enumerate}

In principle, these features could be useful if one were interested in preparing a complete NLO 
parametrization of the process, that includes QCD effects and decay simulation by Monte Carlo.
The expected smoothness of the radiative effects beyond thresholds can be exploited to replace the full
calculation by simple (model dependent) interpolating expressions. 
This is particularly relevant in the SM case where the Sudakov-like parametrization of the process 
is fixed and does not depend on any model parameter, but only on the kinematical cuts.

This remark concludes our presentation of the unpolarized effects. We move 
now to the discussion of the possible polarized observables of the process.

\subsection{Polarized observables}

\subsubsection{Final top asymmetry: one loop effects in $A_{LR}$}

A special property of the $tW^-$ production process is the fact that, in 
principle,  the polarization of the final top quark and/or W boson can be measured. 
This fact, that was first considered in a previous reference~\cite{Gonzalez},
leads to the introduction of new observables, that we 
shall try to list and to discuss in what follows. The first possibility is 
that of measuring the final top polarization. In the process that we are 
considering, the final top can have in principle both helicities, as one can 
see from the expressions of the helicity amplitudes given in Section 3. In 
correspondence to the two possible choices, we shall define two different 
differential cross sections, that we shall define as $d\sigma_{L, R}/ds$, 
that are the analogues of Eq.~(\ref{eq:basic}) where only the contributions 
from the two types of final top have been retained. Plotting these quantities 
at variable $\sqrt{s}$, as we did for the total unpolarized cross section, would lead 
to conclusions that do not much differ from those already given in the 
previous part of this Section: the {\em genuine} SUSY effect is still rather 
modest. Again, the overall MSSM effect is, though, not small. This could be 
seen in the plots of the two distributions,
but from our previous discussion we believe that it might be preferable to 
consider, again, ratios of cross sections. With this aim, we have defined the 
ratio of the integrated cross sections asymmetries, i.e. the quantity

\be
\label{eq:alr}
A_{LR}(s) = \frac{\sigma_L(s)-\sigma_R(s)}{\sigma_L(s)+\sigma_R(s)}{},\ \ \mbox{with}\ \ 
\sigma_{L, R}(s) = \int_{E^2_{threshold}}^{s}\frac{d\sigma_{L, R}}{ds'}\ ds'.
\ee
Figures~(\ref{fig:alrborn},\ref{fig:alr}) shows the values of $A_{LR}$ at variable $\sqrt{s}$. One sees that, considering 
a realistic value e.g. $\sqrt{s} = 500$ GeV, the one-loop effect on the 
asymmetry reaches in all considered SUSY scenarios an absolute value of
slightly less than 1 \%. This number should be compared to 
the realistic overall uncertainty. For the reasons that we have discussed 
previously, we expect essentially a dominance of the purely statistical 
experimental error, whose size will depend on the available integrated 
luminosity. Lacking a dedicated experimental analysis (in preparation), we can 
use as a guidance the preliminary quoted value (for a different single top production 
process, the $t$-channel one) of~\cite{Gonzalez}, that is a (mainly statistical) four percent.

\subsubsection{Final W asymmetry: one loop effects in $A_{TL}$}

In the $tW^-$ production process, the final $W^-$ is real. Therefore one can, in 
principle, measure the primary W polarization. Assuming that this is the 
case, we have defined two quantities that are the analogues of Eq.~(\ref{eq:alr}) and, 
starting from them we have introduced the transverse-longitudinal
asymmetry, defined as
\be
\label{eq:atl}
A_{TL}(s) = \frac{\sigma_{W_T}(s)-\sigma_{W_L}(s)}{\sigma_{W_T}(s)+\sigma_{W_L}(s)}{},\ \ \mbox{with}\ \ 
\sigma_{W_{T, L}}(s) = \int_{E^2_{threshold}}^{s}\frac{d\sigma_{W_{T, L}}}{ds'}\ ds'.
\ee
The numerical values of $A_{TL}$ are shown in Figs.~(\ref{fig:atlborn},\ref{fig:atl}).
In all cases the one-loop effect at the point $\sqrt{s} = 500$ GeV has an absolute 
value of about 0.5 \%. We do not have yet at disposal a suitable experimental analysis for 
this asymmetry, that is in fact being carried on~\cite{forth1}.

\section{Conclusions}
\label{sec:conclusions}

In this paper, we have performed the first complete electroweak one-loop 
analysis of the associated tW production process in the MSSM with mSUGRA 
mechanism of SUSY symmetry breaking. This has been done using a numerical 
program, MINSTREL, that
satisfies the three constraints of cancellation of ultraviolet and infrared 
divergences and of reproduction of asymptotic Sudakov expansions.  We have 
considered various experimental potential observables, both for unpolarized 
and for polarized production. We have found a relatively small genuine SUSY 
effect for the representative SUSY benchmark points that we have selected, and 
a possibly appreciable, mostly of SM origin, overall one-loop effect. We have 
proposed a number of new observables, in general ratios of experimentally 
measurable quantities, that would be essentially free of disturbing 
theoretical QCD and experimental systematic uncertainties. For these 
quantities, the predictions of the MSSM would be rather precise, making them 
appear as possible precision tests of the involved genuine electroweak content 
of the model.The extension of our results to a different MSSM scenario or to 
different SUSY models would be straightforward. The still missing 
corresponding experimental analysis of the various proposed observables of the 
process is being carried on, and will appear in a more complete forthcoming 
paper~\cite{forth1}.

\acknowledgements

We thank C. Carloni Calame, G. Montagna, O. Nicrosini, and F. Piccinini for discussions and suggestions 
about the treatment of QED soft radiation and the related infrared cancellation properties.

\begin{table}
\begin{tabular}{|l|llll|}
\hline
                        &SU1                &SU6        &LS1 &LS2 \\
\hline
$m_0$                    & 70            & 320         & 300              & 300\\
$m_{1/2}$                & 350           & 375         & 150              & 150\\
$A_0$                    & 0             & 0           & -500             & -500\\
$\tan\beta$              & 10            & 50          & 10               &  50\\
$\mu/|\mu|$              & 1             & 1           & 1                &  1\\
$\alpha$                 & -0.110       & -0.0212	 & -0.109	     &      -0.015\\
$M_1$                    & 144.2  	 &     155.8  &  	 60.1	      &    60.6\\
$M_2$                    & 270.1  	  &    291.3   & 	114.8	       &   115.9\\
$\mu$                    & 474.4  	   &   496.6  	 & 329.7	   & 309.3\\
$H^\pm$                  & 534.3  	    &  401.7  	 & 450.4	   & 228.9\\
$H^0$                    & 528.3  	     & 392.5  	 & 442.5	   & 211.1\\
$h^0$                    & 114.6  	     & 115.7  	 & 111.4	   & 110.8\\
$A^0$                    & 527.9  	     & 392.5  	 & 443.4	   & 212.0\\
$\chi^\pm_1$             & 262.8	     & 289.3	 & 108.0	           & 111.1	 \\
$\chi^\pm_2$	         & 495.3	     & 514.8	 & 350.1	 	   & 329.4	 \\
$\chi^0_1$	         & 140.1	     & 153.0	 & 57.38	 	   & 58.92	 \\
$\chi^0_2$	         & 263.1	     & 289.4	 & 108.5	 	   & 111.3	 \\
$\chi^0_3$	         & 479.2	     & 501.0	 & 335.3	 	   & 315.8	 \\
$\chi^0_4$	         & 495.4       &  514.0    	 & 348.7    	   & 326.5    \\
\hline
\end{tabular}
\quad\quad\quad
\begin{tabular}{|l|llll|}
\hline
                        &SU1                &SU6        &LS1 &LS2 \\
\hline
$\widetilde{l}_L$        & 253.3  	     & 412.3  	 & 321.0	   & 321.2\\
$\widetilde{l}_R$        & 157.6  	     & 353.4  	 & 308.7	   & 308.7\\
$\widetilde{\nu}_e$      & 241.0  	     & 404.8  	 & 311.3	   & 311.3\\
$\widetilde{\tau}_L$     & 149.6  	     & 195.8  	 & 297.1	   & 078.1\\
$\widetilde{\tau}_R$     & 256.1        & 399.2  &  	323.8	   & 282.5\\
$\widetilde{\nu}_\tau$   & 240.3        & 362.5  &  	308.4	   & 243.6\\
$\widetilde{u}_L$        & 762.9        & 870.5   & 	459.8	   & 460.2\\
$\widetilde{u}_R$        & 732.9  	 &     840.7  &  	451.9	 &   452.3\\
$\widetilde{d}_L$        & 766.9  	 &     874.0   & 	466.4	  &  467.0\\
$\widetilde{d}_R$        & 730.2  	 &     837.8  	 & 452.8	   & 453.2\\
$\widetilde{t}_L$        & 562.5  	 &     631.5  	 & 213.3	   & 223.6\\
$\widetilde{t}_R$        & 755.8  	 &     796.9  	 & 462.9	   & 431.3\\
$\widetilde{b}_L$        & 701.0  	 &     713.7  	 & 380.6	   & 304.0\\
$\widetilde{b}_R$        & 730.2  	 &     787.6  	 & 449.1	   & 401.7\\
$\theta_\tau$            & 1.366        & 1.133   	 & 1.091	           & 1.117  \\
$\theta_b$               & 0.3619       & 0.7837  	 & 0.184	           & 0.653 \\
$\theta_t$               & 1.070        & 1.050   	 & 1.016             &  0.9313 \\
\hline
\end{tabular}
\caption{
Table of spectra for the various benchmark points. All entries with the dimension of a mass are expressed in GeV. 
The spectra have been computed with the code SUSPECT~\cite{Djouadi:2002ze}.}
\label{tab:spectra}
\end{table}

\newpage

\begin{figure}
\centering
\epsfig{file=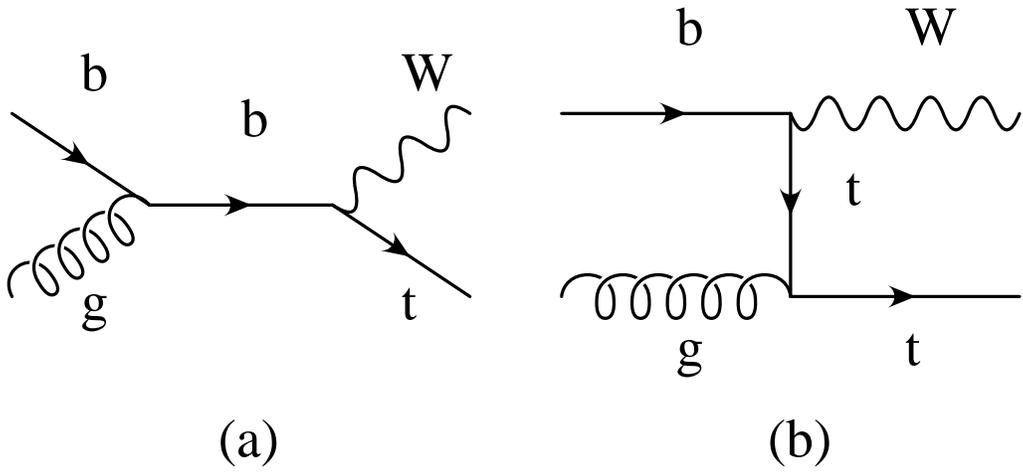, width=14cm, angle=0}
\vspace{1.5cm}
\caption{
Born diagrams for the process $bg\to tW^-$.}
\label{fig:born}
\end{figure}

\newpage

\begin{figure}
\centering
\epsfig{file=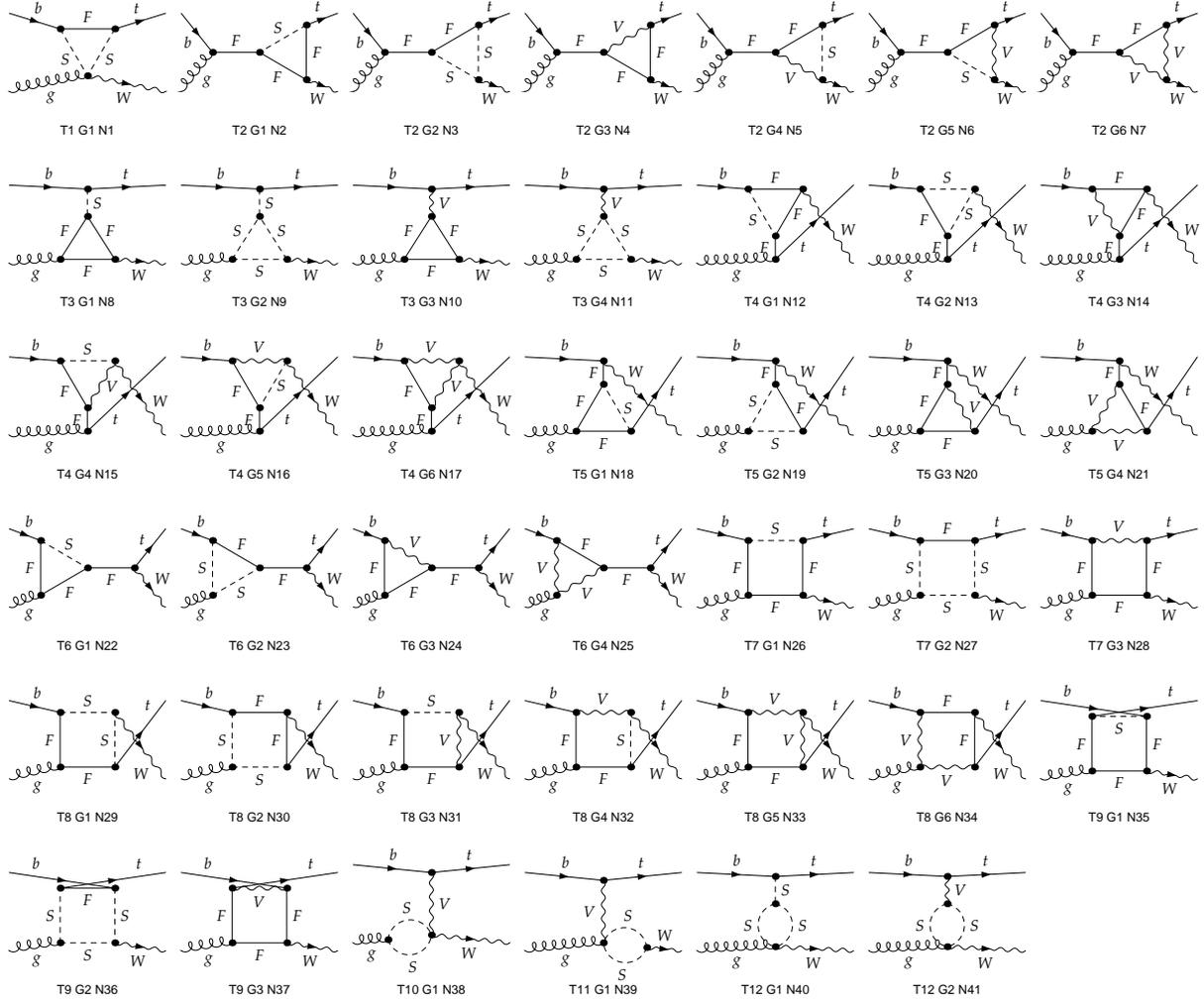, width=16cm, angle=0}
\vspace{1.5cm}
\caption{
Generic diagrams for the process $bg\to tW^-$. We list only the vertex corrections and the box diagrams.
The labels S, F, and V denote generic particles with spin 0, $1/2$, and $1$.
}
\label{fig:generic}
\end{figure}

\newpage

\begin{figure}
\centering
\epsfig{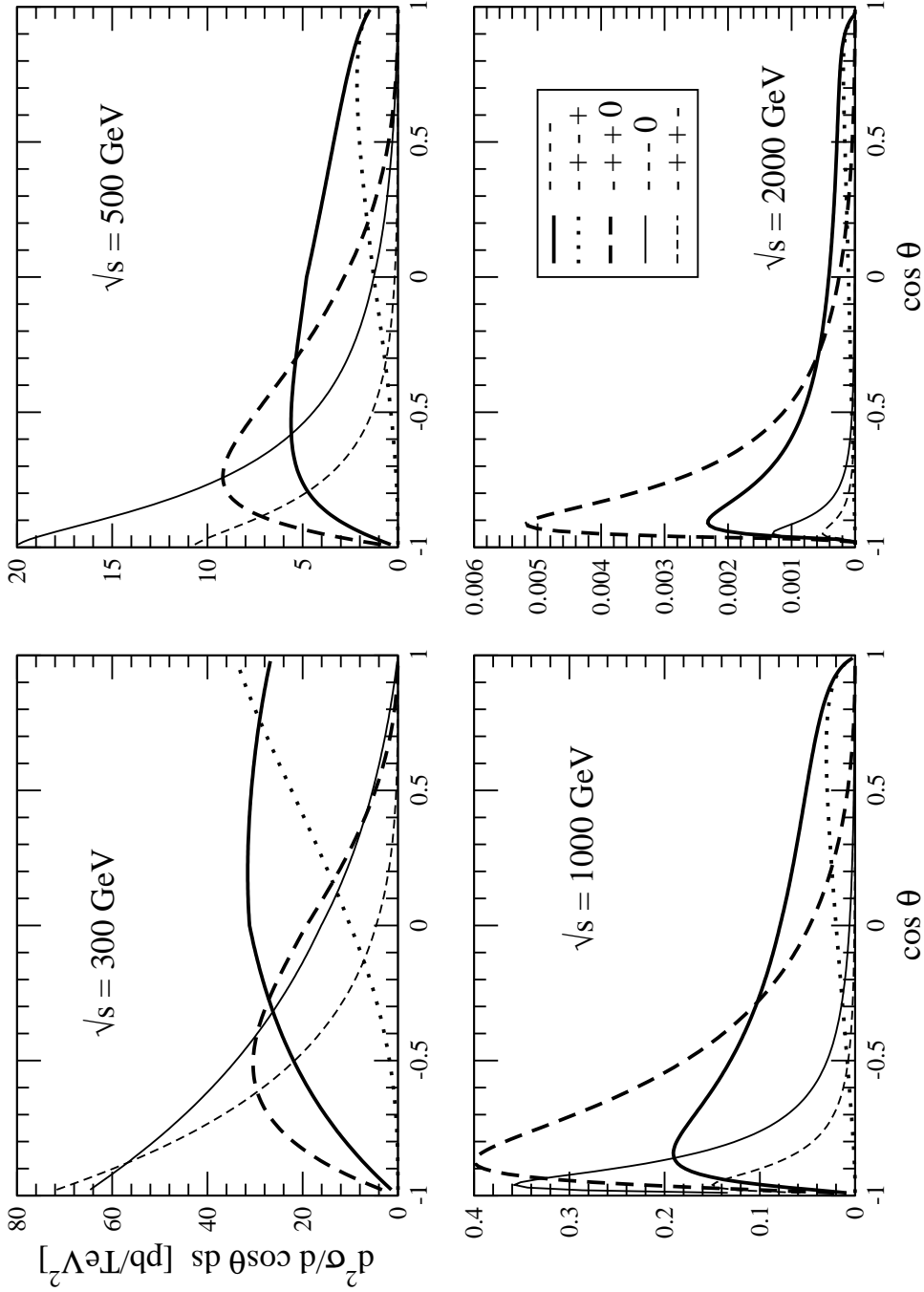}
\vspace{1.5cm}
\caption{
We show the energy and angular dependence of the 5 helicity amplitudes which are 
leading at high energy. Of course, these include the three amplitudes which are not mass suppressed.
In addition, we show the next relevant amplitudes which are the mass suppressed ones $F_{---0}$ and $F_{-++-}$.
An inspection of the Figure shows that below 1 TeV the mass suppression is not effective, especially in the backward
region.
}
\label{fig:diffsigmaborn}
\end{figure}

\newpage

\begin{figure}
\centering
\epsfig{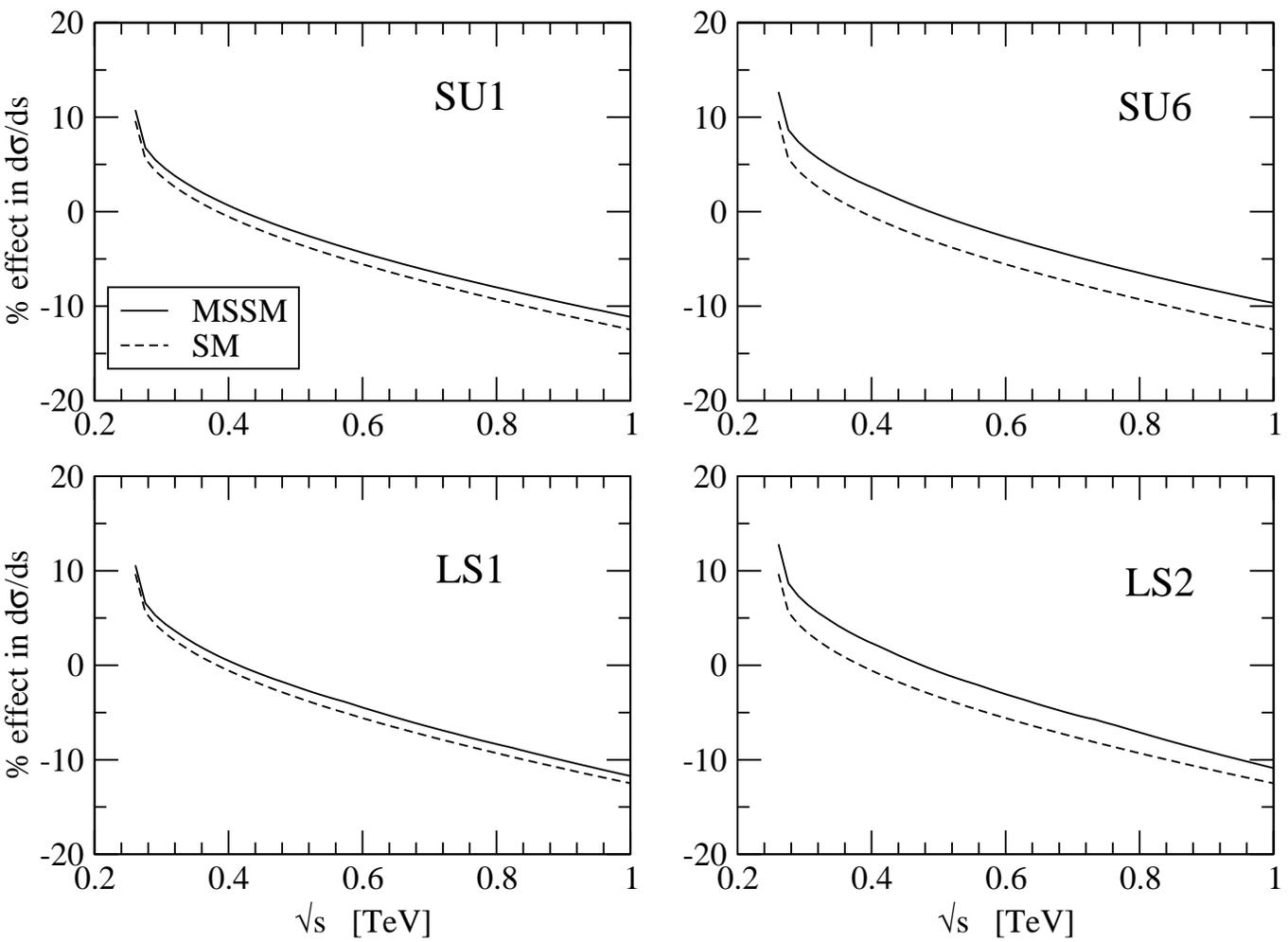}
\vspace{1.5cm}
\caption{
MSSM and SM one loop effect in the distribution $d\sigma/ds$ in the four considered scenarios
SU1, SU6, LS1, LS2.
}
\label{fig:sigma}
\end{figure}

\newpage

\begin{figure}
\centering
\epsfig{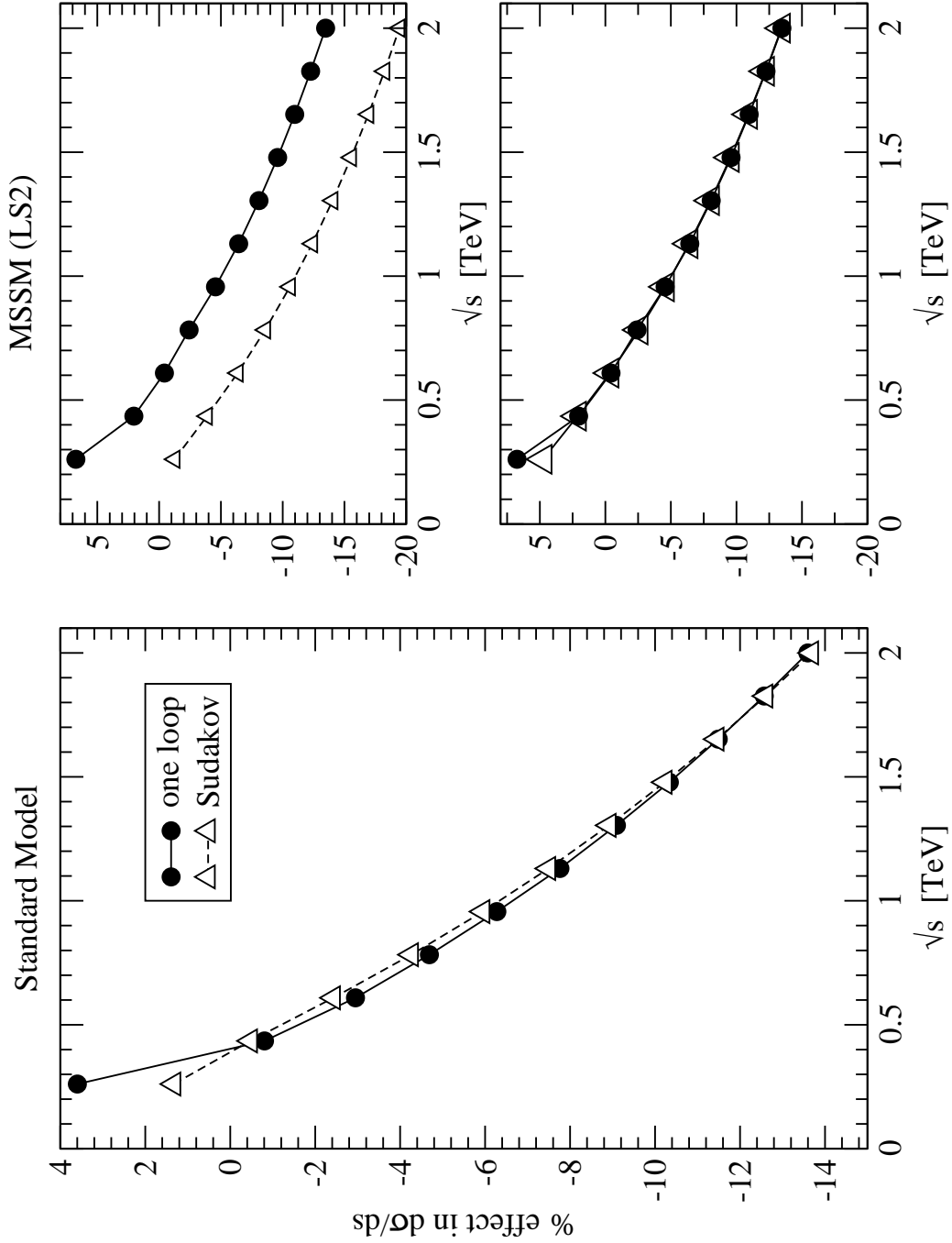}
\vspace{1.5cm}
\caption{
Comparison between the full one loop and Sudakov  calculations of the effect in the distribution $d\sigma/ds$.
The left panel shows the Standard Model case. The right panel shows the LS2 MSSM scenario, which is the lightest considered.
In the upper figure we show the two curves, whereas in the lower panel, we have shifted the Sudakov effect by a constant +6\%.
For purposes of comparison, we have switched off QED radiation and set $M_\gamma = M_Z$.
}
\label{fig:sudakov}
\end{figure}

%\newpage
%
%\begin{figure}
%\centering
%\epsfig{file=PaperSigmaDelta.eps, width=18cm, angle=90}
%\vspace{1.5cm}
%\caption{
%Effective Sudakov-like parametrization of the one loop effect in the distribution $d\sigma/ds$ in the four considered scenarios
%SU1, SU6, LS1, LS2. We show the MSSM case only. The SM one is summarized in Tab.~(\ref{tab:delta}) and has a similar quality.
%}
%\label{fig:delta}
%\end{figure}

\newpage

\begin{figure}
\centering
\epsfig{file=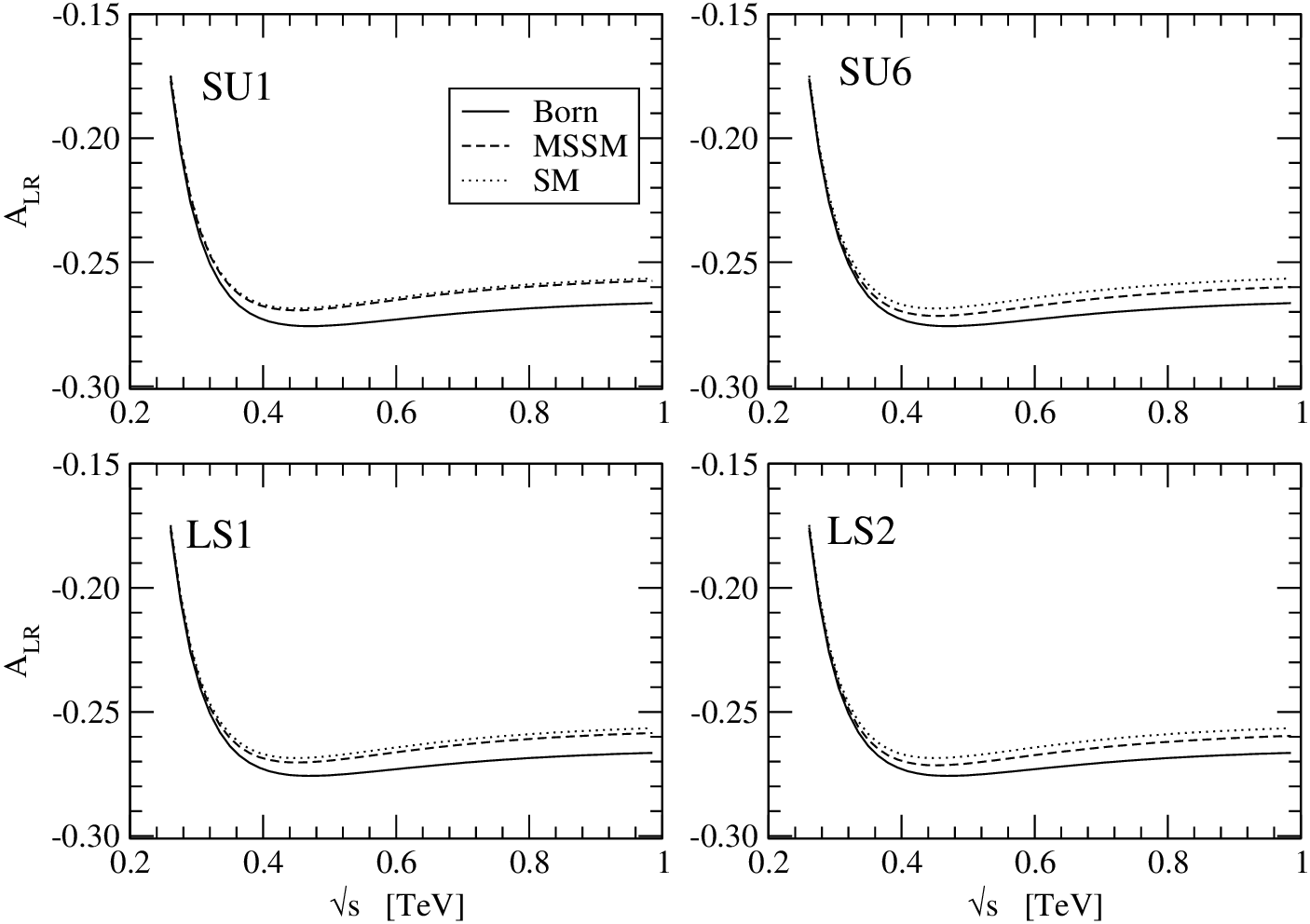, width=18cm, angle=90}
\vspace{1.5cm}
\caption{
Born value and MSSM/SM one loop effect in the asymmetry $A_{LR}$ in the four considered scenarios
SU1, SU6, LS1, LS2. The cross sections entering $A_{LR}$ are integrated from threshold up to $\sqrt{s}$.
}
\label{fig:alrborn}
\end{figure}

\newpage

\begin{figure}
\centering
\epsfig{file=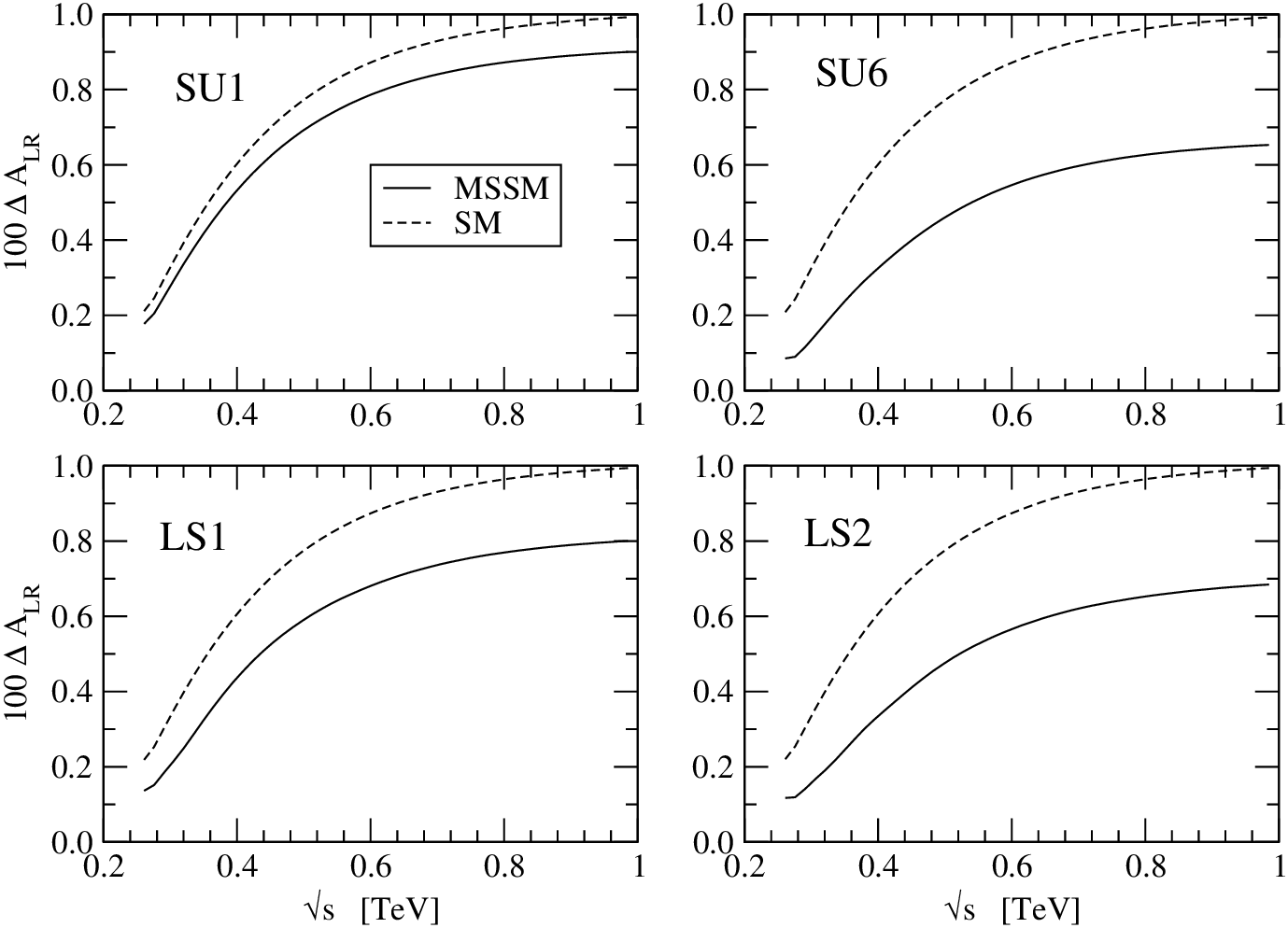, width=18cm, angle=90}
\vspace{1.5cm}
\caption{
Percentual MSSM and SM one loop effect in the asymmetry $A_{LR}$ in the four considered scenarios
SU1, SU6, LS1, LS2. The cross sections entering $A_{LR}$ are integrated from threshold up to $\sqrt{s}$.
}
\label{fig:alr}
\end{figure}

\newpage

\begin{figure}
\centering
\epsfig{file=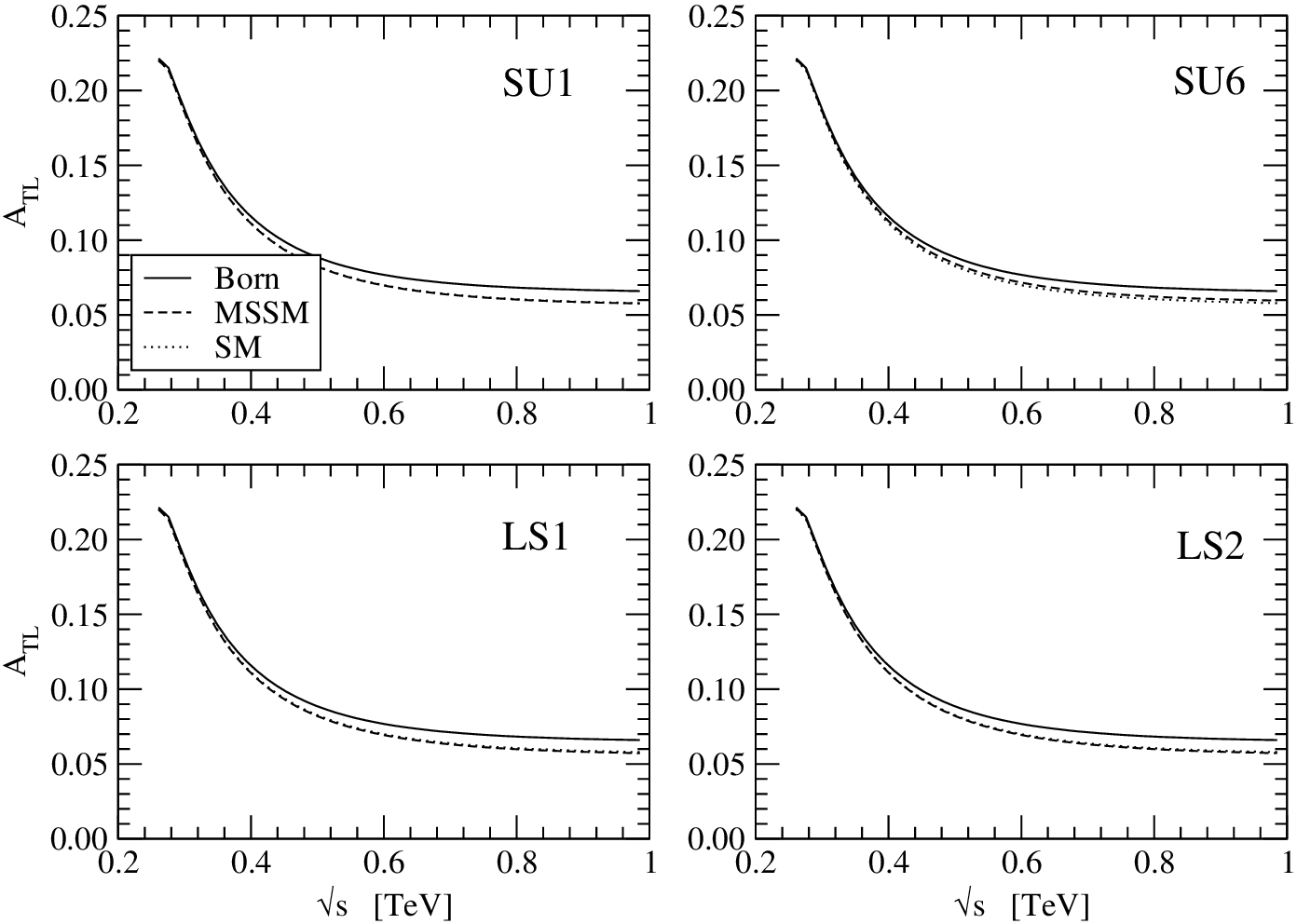, width=18cm, angle=90}
\vspace{1.5cm}
\caption{
Born value and MSSM/SM one loop effect in the asymmetry $A_{TL}$ in the four considered scenarios
SU1, SU6, LS1, LS2. The cross sections entering $A_{TL}$ are integrated from threshold up to $\sqrt{s}$.
}
\label{fig:atlborn}
\end{figure}

\newpage

\begin{figure}
\centering
\epsfig{file=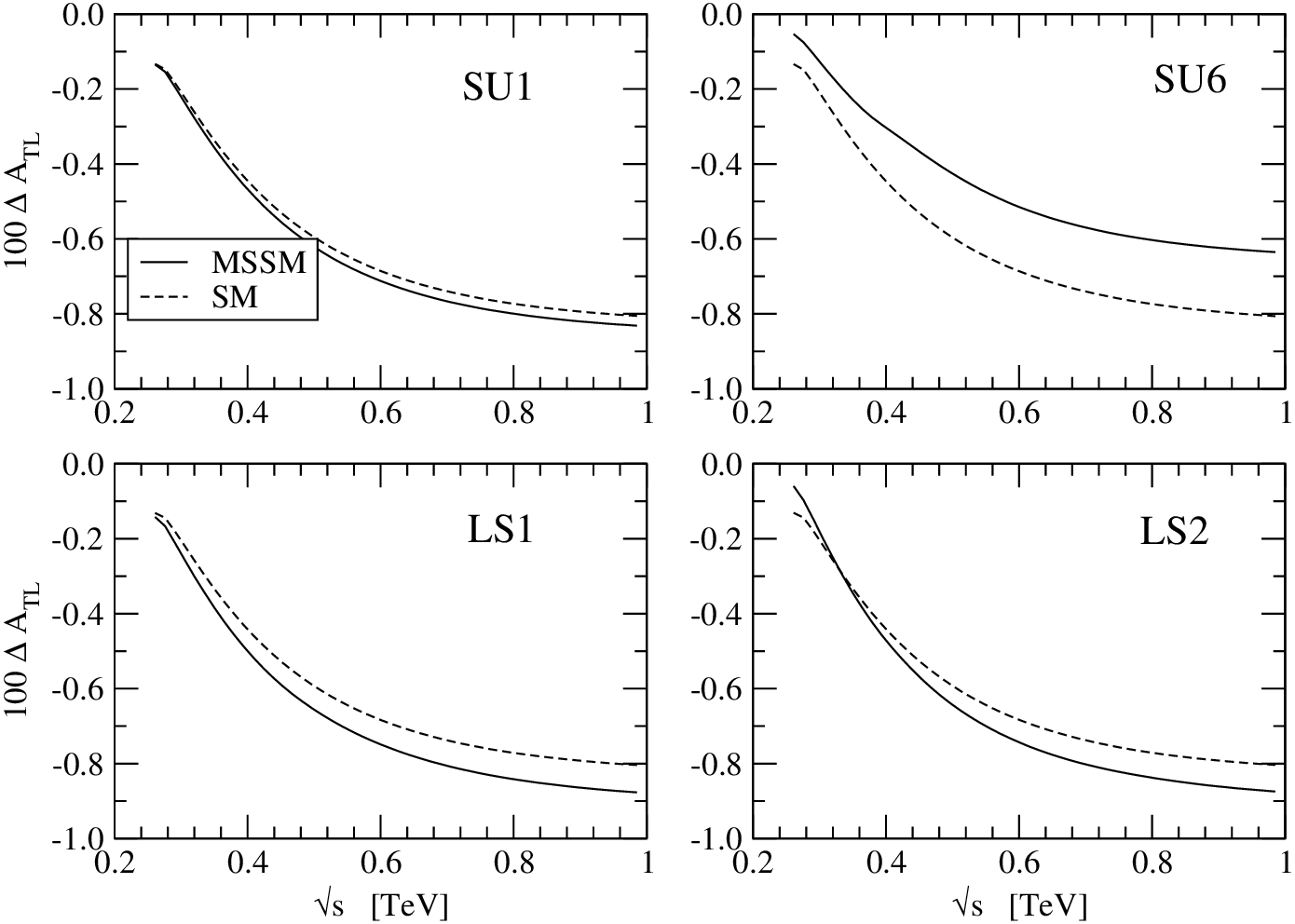, width=18cm, angle=90}
\vspace{1.5cm}
\caption{
Percentual MSSM and SM one loop effect in the asymmetry $A_{TL}$ in the four considered scenarios
SU1, SU6, LS1, LS2. The cross sections entering $A_{TL}$ are integrated from threshold up to  $\sqrt{s}$.
}
\label{fig:atl}
\end{figure}

\end{document}